\documentclass[a4paper,10pt]{amsart}

\usepackage{amsmath,amssymb,amsthm,a4wide,hyperref,xcolor}

\usepackage{a4wide}
\usepackage{amsmath,amssymb,amsthm}
\usepackage[pdftex]{graphicx}
\usepackage{graphics}
\usepackage{tikz}
\usepackage{color}

\newcommand\NN {\mathbb{N}}

\title[Local extrema in Quantum Chaos]{On the Distribution of Local extrema in Quantum Chaos}
\author[Florian Pausinger]{Florian Pausinger$^{\dagger}$}\thanks{$^{\dagger}$ IST Austria, Am Campus 1, 3400 Klosterneuburg, Austria, e-mail: \textsc{florian.pausinger@ist.ac.at}}

\author[Stefan Steinerberger]{Stefan Steinerberger$^{\dagger \dagger}$}\thanks{$^{\dagger \dagger}$ Department of Mathematics, Yale University, 10 Hillhouse Avenue, New Haven, CT 06511, USA e-mail: \textsc{stefan.steinerberger@yale.edu} (corresponding author)}


\begin{document}
\begin{abstract} We numerically investigate the distribution of extrema of 'chaotic' Laplacian eigenfunctions on two-dimensional manifolds. Our contribution is two-fold:
(a) we count extrema on grid graphs with a small number of randomly added edges and show the behavior to coincide with the 1957 prediction of Longuet-Higgins
for the continuous case and
(b) compute the regularity of their spatial distribution using \textit{discrepancy}, which is a classical measure from the theory of Monte Carlo integration. The first part
suggests that grid graphs with randomly added edges should behave like two-dimensional surfaces with ergodic geodesic flow; in the second part we show that
the extrema are more regularly distributed in space than the grid $\mathbb{Z}^2$.
\end{abstract}
\maketitle


\section{Introduction}
\subsection{Quantum Chaos.} Quantum Chaos is concerned with the behavior of high-frequency Laplacian eigenfunctions 
$$ -\Delta u = E u \qquad \mbox{on compact manifolds}~(M,g)$$
and their seemingly chaotic properties. Apart from highly particular cases
which are usually characterized by completely integrable behavior of the geodesic flow, 
these eigenfunctions will appear to be somewhat 'random'. Indeed, should the behavior be not
chaotic, then usually any small perturbation of the geometry of the domain will induce chaotic
behavior: randomness is the generic case. It is of great interest to try to understand this randomness
by specifying arising invariants. 
\begin{figure}[h!]
\begin{minipage}[t]{0.49\columnwidth}%
\begin{center}
\includegraphics[width = 6cm]{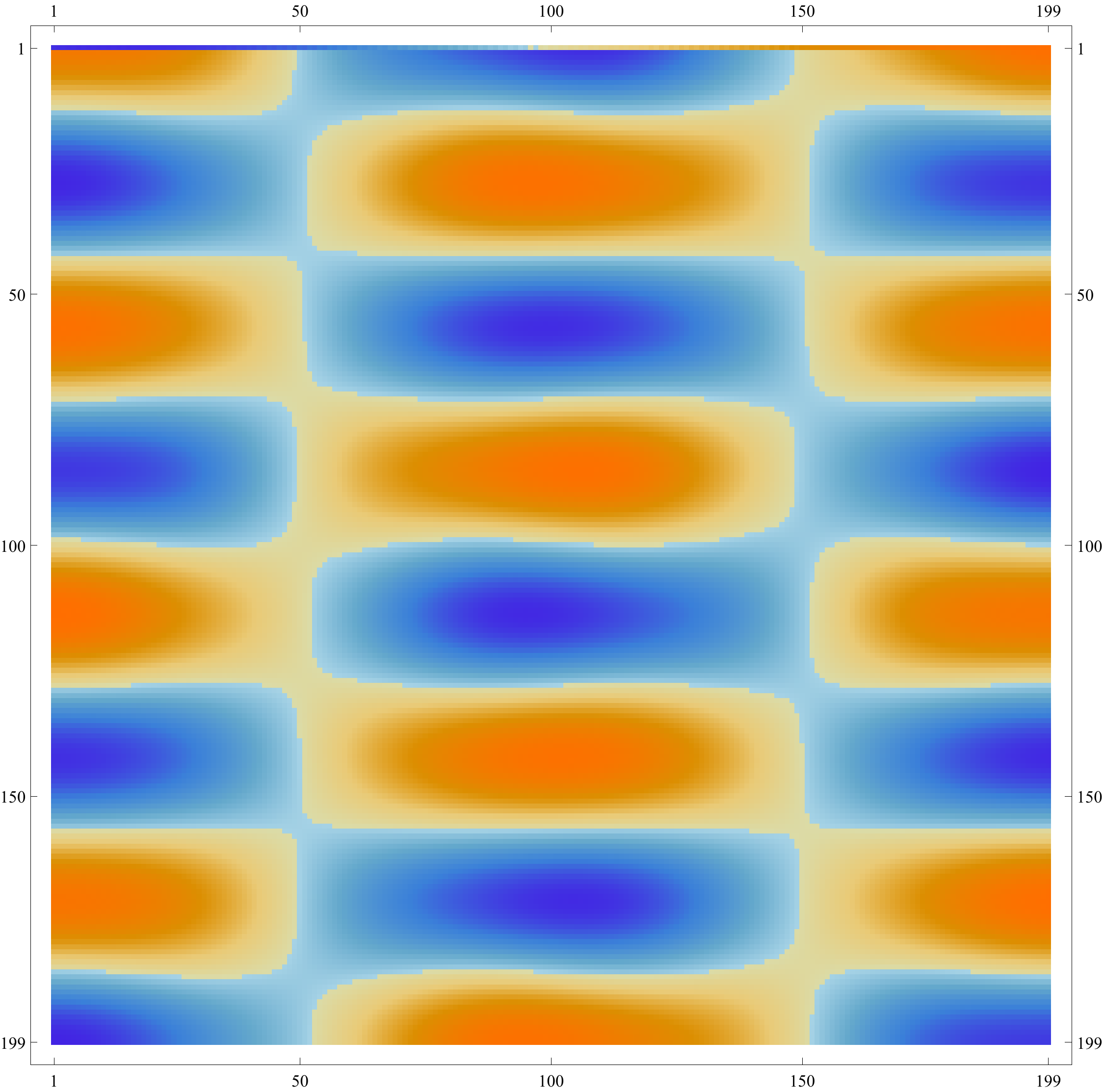}
\end{center}
\end{minipage}%
\begin{minipage}[t]{0.49\columnwidth}%
\begin{center}
\includegraphics[width = 6cm]{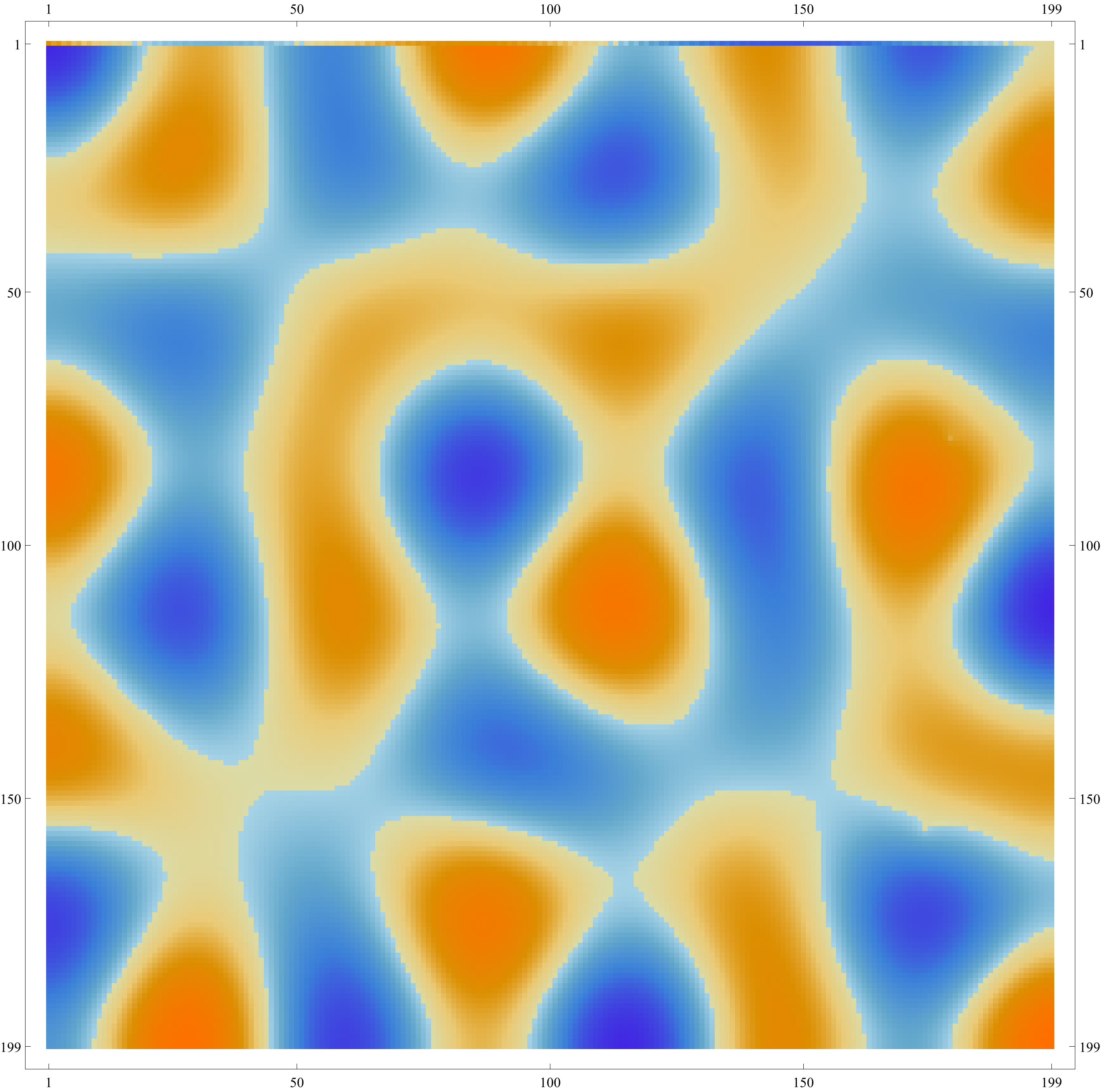}
\end{center}
\end{minipage}
\caption{An eigenfunction of the Laplacian with Neumann conditions once on
$[0,1]^2$ (left) and once on a small perturbation of $[0,1]^2$ (right, the perturbation is not visible).}
\end{figure}

Some central questions of quantum chaos are 
\begin{enumerate}
\item whether (and under which conditions on the geometry of
the manifold) the $L^2-$mass of the eigenfunctions tends towards uniform distribution --
recent spectacular breakthroughs due to Anantharaman \cite{ana} and Lindenstrauss \cite{lind}.
\item whether most eigenfunctions behave like 'random waves', i.e. whether for example
$$ \frac{\|u_k\|_{L^{\infty}}}{\|u_k\|_{L^{2}}} \lesssim (\log{k})^{\frac{1}{2}} \qquad
\mbox{with high probability \cite{au}}$$
\item how many nodal domains there are (see \cite{blum, bog, bourgain2} for the random wave model
and \cite{bourgain, stein} for deterministic bounds) and how their volume is distributed (see e.g. \cite{toth}).
\end{enumerate}
The number of nodal domains has received particular interest: in a highly influental paper by Blum, Gnutzmann \& Smilansky \cite{blum}, a universality statement for the number of nodal domains has been conjectured and numerically investigated: a generic Laplacian eigenfunction associated to the $k-$th eigenvalue seems to have $\sim 0.06k$ nodal domains. Bogomolny \& Schmit \cite{bog} have worked out a percolation model simulating eigenfunctions in which the observation of Blum, Gnutzmann \& Smilansky is confirmed: their model predicts that the number of nodal domains of the $k-$th eigenfunction is distributed with 
$$\frac{3\sqrt{3}-5}{\pi}k \sim 0.06k \quad \mbox{mean and a variance of} \quad
\left(\frac{18}{\pi^2} +\frac{4\sqrt{3}}{\pi} - \frac{25}{2\pi}\right)k \sim 0.05k.$$
It is not yet understood to what extent these numbers are precise outside the model (recent numerical work of Konrad \cite{konrad} suggests the mean to be $\sim 4\%$ smaller), however, they are certainly very good approximations.

\subsection{Chaotic eigenfunctions, local extrema and finite graphs.} 
We are interested in the distribution of the local extrema of a Laplacian eigenfunction on a two-dimensional smooth surface with non-integrable geodesic flow. 
A cornerstone of existing conjectures is the random wave heuristic, which asserts that for all practical purposes a Laplacian
eigenfunction should behave like a superposition of random plane waves
$$ \psi(\textbf{r}) = \sum_n{a_k\cos{(\left\langle \textbf{k}_n, \textbf{r}\right\rangle - \phi_n})},$$
where $a_k, \phi_k$ are random reals and $\textbf{k}$ is a randomly chosen direction normalized to $\|\textbf{k}\| = \sqrt{E}$, where $E$ is the energy/eigenvalue.
Longuet-Higgins \cite{lon} studied this heuristic in a pioneering 1957 paper, which suggests that the $n-$th Laplacian eigenfunction on a compact two-dimensional surface
should have $\sim n/\sqrt{3}$ extrema. The random wave approximation is of fundamental importance as its framework allows for precise computations while precise mathematical
result seem still out of reach: for example, one would expect (see e.g. Yau \cite{yau}) that the nodal length has a $(n-1)-$dimensional Hausdorff measure of size $\sim \sqrt{E}$ while the currently
 best rigorous results in dimensions $\geq 4$ \cite{cm,sz,ss} do not even rule out the possibility that the nodal length might tend to 0 as $E \rightarrow \infty$.\\

It is natural to try simpler examples; a prime candidate is a reduction to finite graphs $G = (V,E)$.
Given a finite, simple, connected Graph $G = (V, E)$ the natural analogue of the Laplacian is the discrete Graph-Laplacian (see e.g. \cite{chung}) given by a $|V| \times |V|$-matrix $L$ with entries
$$ L_{ij} = \begin{cases} 1 \qquad &\mbox{if}~i=j \\
-(d_i d_j)^{-\frac{1}{2}} \qquad &\mbox{otherwise,}\end{cases}$$
where $d_i$ is the degree of the vertex $i$. The first few eigenvalues/eigenvectors of the matrix
will then approximate the first few eigenvalues/eigenfunctions of the Laplacian with Neumann
boundary condition (for details we refer to \cite{singer}). It is not difficult to see that as the graph increases in size, it may be used to approximate the given geometry
to any arbitrary degree of accuracy; however, counting nodal domains on graphs is rather difficult. Clearly, if two vertices $u,v$ are joined by an edge e $u \sim_{e} v$ and the eigenfunction
satisfies $f(u)f(v) < 0$, one would say that the edge crosses the nodal domain -- the lack of continuity does not allow for an immediate transfer of the definition from the
continuous case. Indeed, nodal domains of eigenfunctions on graphs are an ongoing field of great interest
(see e.g. Davies, Gladwell, Leydold \& Stadler \cite{dav}, Dekel, Lee \& Linial \cite{dek} or the survey \cite{band}) but difficulty
in transferring even very classical theorems from the continuous to the discrete setting pose a difficulty. In contrast, an extremum of a function is topologically simpler than 
that of a nodal domain and more easily generalized to the setting of a graph.

\begin{quote}
\textbf{Contribution 1.}  Our first contribution is that
grid graphs with a small number of added edges behave like continuous surfaces with respect to the number of extrema of eigenfunctions and recover Longuet-Higgins
prediction. We also computed random wave approximations on both $\mathbb{T}^2$ and $\mathbb{S}^2$ to allow for comparison. 
\end{quote}
Using the fact that grid graphs with a small number of randomly added edges seem to provide a second way (the other being the random wave approximation) to create 
essentially 'chaotic' behavior, we use both ways to try to understand the way the extrema are distributed in space. We hasten to emphasize that graphs have, of course, been
used by many people to describe chaotic behavior (see e.g. a paper of Smilansky \cite{smigraph}, where $d-$regular graphs are employed); one possible advantage 
of using grid graphs with a random number of edges is their simplicity (the downside being, of course, that expander graphs, to give just one example, come with
many additional properties which simplify a rigorous analysis).

\begin{figure}[h!]
\begin{minipage}[t]{0.49\columnwidth}%
\begin{center}
\includegraphics[width = 6.8cm]{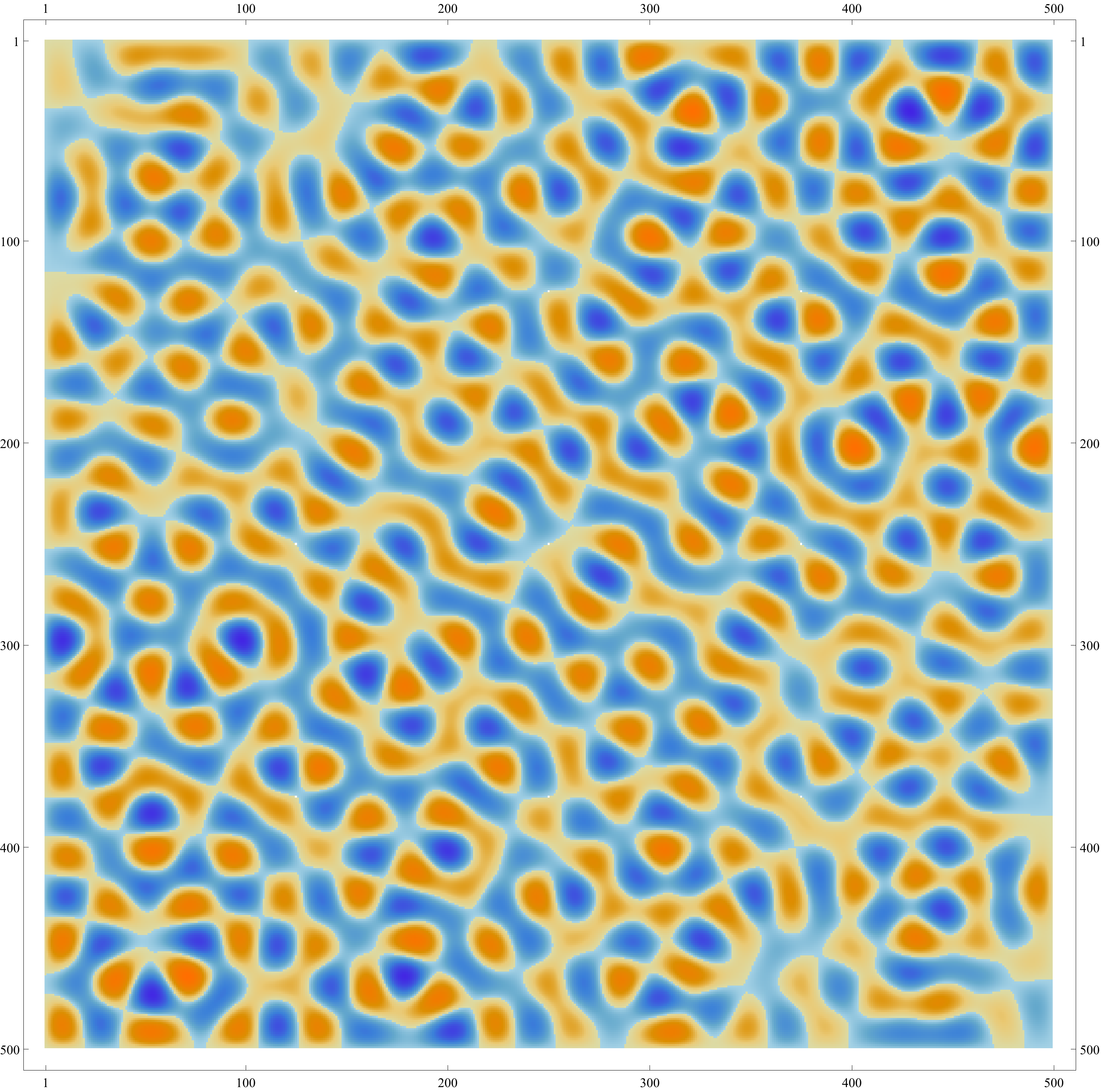}
\end{center}
\end{minipage}%
\begin{minipage}[t]{0.49\columnwidth}%
\begin{center}
\includegraphics[width = 6.8cm]{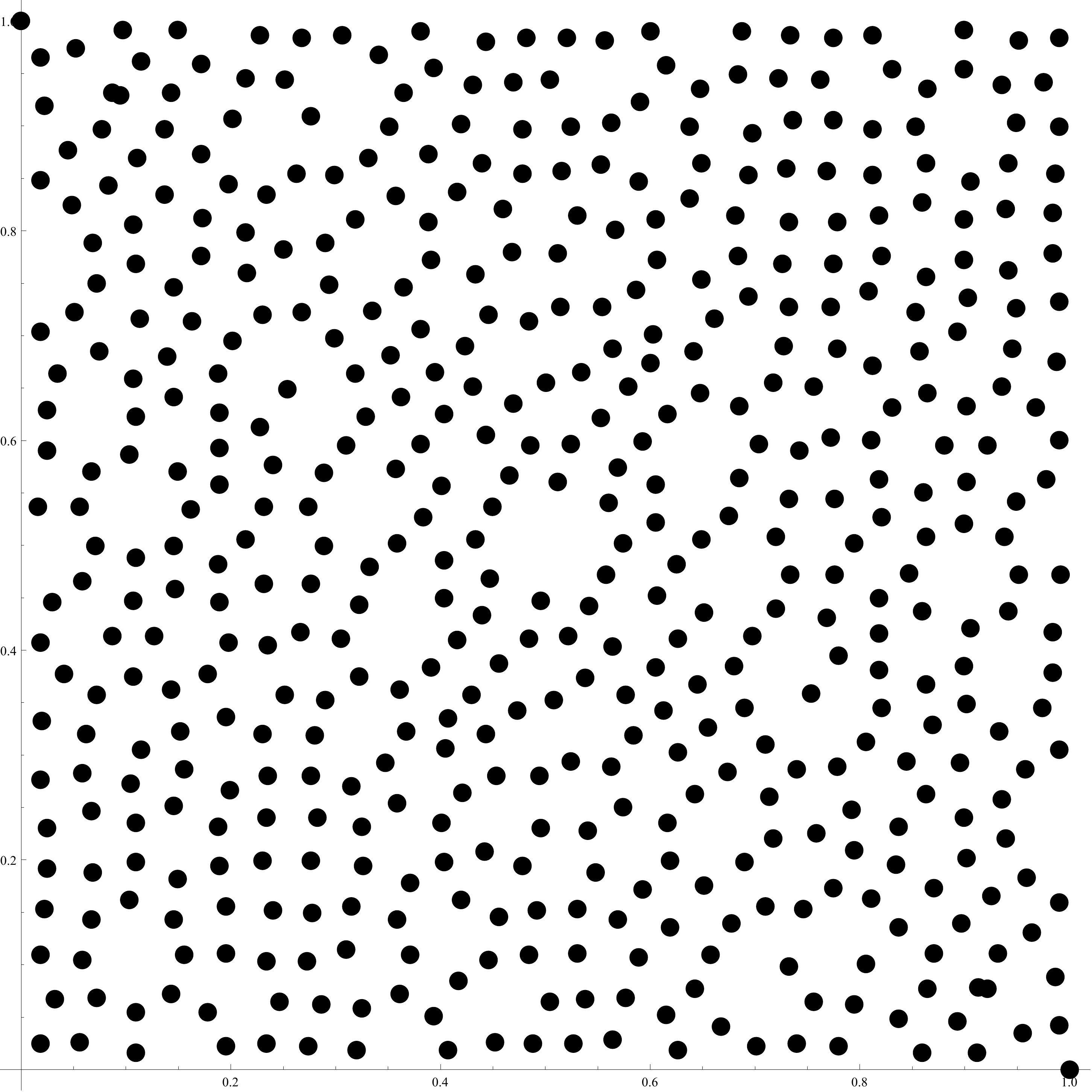}
\end{center}
\end{minipage}
\caption{A 'chaotic' eigenfunction of the Laplacian on $\mathbb{T}^2$ and its local extrema.}
\end{figure}

When studying the distribution of local extrema in space, we use discrepancy as a quantitative measure of regularity. Discrepancy is \textit{the} standard measure in theory of uniform distribution (c.f. classical books of Niederreiter \cite{nied} and Drmota \& Tichy \cite{drm})
and has further applications in the theory of quasi-Monte Carlo integration. A point set with a small discrepancy is thus both well distributed from an abstract point of view as well as very suitable
for numerical integration of a function with controlled oscillation. 
\begin{quote}
\textbf{Contribution 2.} Extrema of chaotic Laplacian eigenfunctions are more regularly distributed with respect to discrepancy than the (suitably rescaled) classical grid $\mathbb{Z}^2$. In particular, they
are better suited for numerical integration than the extrema of non-chaotic eigenfunctions (the extrema of the eigenfunction $\sin{n \pi x}\sin{n \pi y}$ of $-\Delta$ on $[0,1]^2$ are a translation of
a rescaling of the grid $\mathbb{Z}^2$).
\end{quote}

The rest of the paper is structured as follows: in Section 2 we describe our heuristic reasoning for why to employ grid graphs with randomly added edges to create quantum chaos, Section 3
shows that these graphs are able to reproduce the prediction of Longuet-Higgins on the number of local extrema, Section 4 introduces the discrepancy and 
describes the numerical results about the spatial distribution of Laplacian eigenfunction using both random wave models and the grid graphs with randomly added edges; technical comments and
details about the implementation are given in the final section.


\section{Generating quantum chaos}
\subsection{Motivation} Our approach was motivated by the following basic observation: the set of
extrema of the function $\sin{(n\pi x)}$, which is a solution to the eigenvalue problem
$$ -\Delta u = n^2 \pi^2 u \qquad \mbox{on}~[0,1]^2,$$
is extremely regular.
\begin{center}
\begin{figure}[h!]
\begin{tikzpicture}[xscale=2.5]
  \draw[thick] (0,0) -- (3.1415,0);
\draw [fill] (3.1415/10,0) ellipse (0.02cm and 0.05cm);
\draw [fill] (3*3.1415/10,0) ellipse (0.02cm and 0.05cm);
\draw [fill] (5*3.1415/10,0) ellipse (0.02cm and 0.05cm);
\draw [fill] (7*3.1415/10,0) ellipse (0.02cm and 0.05cm);
\draw [fill] (9*3.1415/10,0) ellipse (0.02cm and 0.05cm);
  \draw[domain=0:3.14159265,smooth,variable=\x,black, thick] plot (\x,{sin(5 * \x r)});
\end{tikzpicture}
\caption{The location of the extrema of $\sin{(5\pi x)}$ on the unit interval.}
\end{figure}
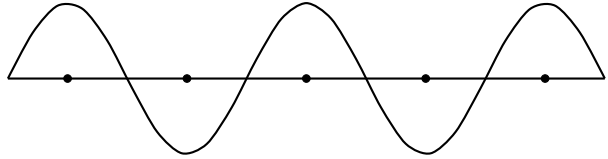
\end{center}
This is not altogether surprising: these eigenfunctions of the Laplacians may be alternatively
described as minima of the Rayleigh quotient
$$ \inf_{f \in H^{1}_0(\Omega)}{\frac{\int_{\Omega}{|\nabla u|^2dx}}{\int_{\Omega}{u^2dx}}}
\qquad \mbox{with}~f~\mbox{being orthogonal on previous eigenfunctions.}$$
It is therefore not strange that two different extrema should be at a certain distance from each other: a maximum and a minimum too close together would induce a large gradient between them. Let us
now consider the obvious generalization (with Dirichlet boundary conditions)
$$ -\Delta u = \lambda u \qquad \mbox{on the domain}~[0,1]^2.$$
It is well-known that the solutions are of the form
$$ \left\{\sin{(n\pi x)}\sin{(m \pi x)}: m,n \in \mathbb{N} \wedge \pi^2(m^2+n^2) = \lambda\right\}.$$
The extrema have a very regular distribution: if $m=n$, they are merely a rescaling and translation of the grid $\mathbb{Z}^2$ while for $m \neq n$ there is an additional 
affine transformation -- there is no chaos. The multiplicative splitting of the eigenfunction is induced by the geometry of the domain and a \textit{highly} unstable 
phenomenon: any generic small perturbation of the domain will destroy that property. A natural question is how the introduction of chaos will affect the regularity: 
interestingly, we will see that chaos actually improves it.

\subsection{Generating Chaos.} Our approach is based on the idea of identifying pairs of points on the unit square $[0,1]^2$: this corresponds to introducing a small 'wormhole' connecting two points
 that were initially far away.
\begin{center}
\begin{figure}[h!]
\begin{tikzpicture}[scale=3]
  \draw[ultra thick] (0,0) -- (1,0);
  \draw[ultra thick] (0,0) -- (0.5,0.5);
  \draw[ultra thick] (1,0) -- (1.5,0.5);
  \draw[ultra thick] (0.5,0.5) -- (1.5,0.5);

  \draw[ultra thick] (2,0) -- (3,0);
  \draw[ultra thick] (2,0) -- (2.5,0.5);
  \draw[ultra thick] (3,0) -- (3.5,0.5);
\draw [ultra thick] (2.4,0.2) to [out=0,in=180] (2.8,0.7) to [out=0,in=180] (3.1,0.3);
\draw [ultra thick] (2.8,0.2) to [out=180,in=180] (2.8,0.7);
\draw [ultra thick] (2.8,0.3) to [out=0,in=0] (2.8,0.7);
  \draw[ultra thick] (2.5,0.5) -- (2.6,0.5);
  \draw[ultra thick] (2.7,0.5) -- (2.88,0.5);
  \draw[ultra thick] (2.99,0.5) -- (3.5,0.5);
\end{tikzpicture}
\caption{The flat square and the flat square with two points connected by a small wormhole.}
\end{figure}
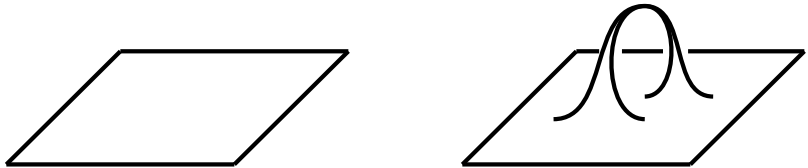
\end{center}
\vspace{-10pt}
To simplify matters, we will not give a geometrically precise definition of the nature of the identification
but instead approximate the unit square $[0,1]^2$ by a finite graph and then simply describe the
modification being carried out on the finite graph. It follows immediately from work of Colin de Verdiere
\cite{cdv} that any such graph can conversely be approximated by a manifold which implies that 
performing the surgery directly on the graph is not a restriction to the discrete setting. We will be working on the grid-graph $G_n$ defined on the vertex set
$$V(G_n) = \left\{1, \dots, n\right\}^2,$$
where two vertices are connected by an edge if and only if
$$(a,b) \sim_E (c,d) :\Leftrightarrow |a-c| + |b-d| = 1.$$
We will then induce a slight perturbation of the geometry by adding a few extra edges (which correspond to the 'wormholes'): interestingly, the results we obtain depend slightly on the number of added edges (where a rather small number, say 5 extra edges for a $n=100$, suffices) but not really \textit{where}
the edges are added -- we will actually add them at random. 

\begin{center}
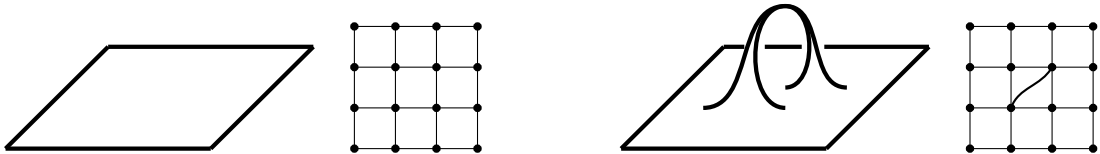
\begin{figure}[h]
\begin{tikzpicture}[scale=2.7]
  \draw[ultra thick] (0,0) -- (1,0);
  \draw[ultra thick] (0,0) -- (0.5,0.5);
  \draw[ultra thick] (1,0) -- (1.5,0.5);
  \draw[ultra thick] (0.5,0.5) -- (1.5,0.5);
\draw [fill, ultra thick] (1.7,0) circle [radius=0.01];
\draw [fill, ultra thick] (1.9,0) circle [radius=0.01];
\draw [fill, ultra thick] (2.1,0) circle [radius=0.01];
\draw [fill, ultra thick] (2.3,0) circle [radius=0.01];
\draw [fill, ultra thick] (1.7,0.2) circle [radius=0.01];
\draw [fill, ultra thick] (1.9,0.2) circle [radius=0.01];
\draw [fill, ultra thick] (2.1,0.2) circle [radius=0.01];
\draw [fill, ultra thick] (2.3,0.2) circle [radius=0.01];
\draw [fill, ultra thick] (1.7,0.4) circle [radius=0.01];
\draw [fill, ultra thick] (1.9,0.4) circle [radius=0.01];
\draw [fill, ultra thick] (2.1,0.4) circle [radius=0.01];
\draw [fill, ultra thick] (2.3,0.4) circle [radius=0.01];
\draw [fill, ultra thick] (1.7,0.6) circle [radius=0.01];
\draw [fill, ultra thick] (1.9,0.6) circle [radius=0.01];
\draw [fill, ultra thick] (2.1,0.6) circle [radius=0.01];
\draw [fill, ultra thick] (2.3,0.6) circle [radius=0.01];
  \draw (1.7,0) -- (2.3,0);
  \draw (1.7,0.2) -- (2.3,0.2);
  \draw (1.7,0.4) -- (2.3,0.4);
  \draw (1.7,0.6) -- (2.3,0.6);
  \draw (1.7,0) -- (1.7,0.6);
  \draw (1.9,0) -- (1.9,0.6);
  \draw (2.1,0) -- (2.1,0.6);
  \draw (2.3,0) -- (2.3,0.6);
  \draw[ultra thick] (3,0) -- (4,0);
  \draw[ultra thick] (3,0) -- (3.5,0.5);
  \draw[ultra thick] (4,0) -- (4.5,0.5);
\draw [ultra thick] (3.4,0.2) to [out=0,in=180] (3.8,0.7) to [out=0,in=180] (4.1,0.3);
\draw [ultra thick] (3.8,0.2) to [out=180,in=180] (3.8,0.7);
\draw [ultra thick] (3.8,0.3) to [out=0,in=0] (3.8,0.7);
  \draw[ultra thick] (3.5,0.5) -- (3.6,0.5);
  \draw[ultra thick] (3.7,0.5) -- (3.88,0.5);
  \draw[ultra thick] (3.99,0.5) -- (4.5,0.5);
  \draw (4.7,0) -- (5.3,0);
  \draw (4.7,0.2) -- (5.3,0.2);
  \draw (4.7,0.4) -- (5.3,0.4);
  \draw (4.7,0.6) -- (5.3,0.6);
  \draw (4.7,0) -- (4.7,0.6);
  \draw (4.9,0) -- (4.9,0.6);
  \draw (5.1,0) -- (5.1,0.6);
  \draw (5.3,0) -- (5.3,0.6);
\draw [fill, ultra thick] (4.7,0) circle [radius=0.01];
\draw [fill, ultra thick] (4.9,0) circle [radius=0.01];
\draw [fill, ultra thick] (5.1,0) circle [radius=0.01];
\draw [fill, ultra thick] (5.3,0) circle [radius=0.01];
\draw [fill, ultra thick] (4.7,0.2) circle [radius=0.01];
\draw [fill, ultra thick] (4.9,0.2) circle [radius=0.01];
\draw [fill, ultra thick] (5.1,0.2) circle [radius=0.01];
\draw [fill, ultra thick] (5.3,0.2) circle [radius=0.01];
\draw [fill, ultra thick] (4.7,0.4) circle [radius=0.01];
\draw [fill, ultra thick] (4.9,0.4) circle [radius=0.01];
\draw [fill, ultra thick] (5.1,0.4) circle [radius=0.01];
\draw [fill, ultra thick] (5.3,0.4) circle [radius=0.01];
\draw [fill, ultra thick] (4.7,0.6) circle [radius=0.01];
\draw [fill, ultra thick] (4.9,0.6) circle [radius=0.01];
\draw [fill, ultra thick] (5.1,0.6) circle [radius=0.01];
\draw [fill, ultra thick] (5.3,0.6) circle [radius=0.01];
\draw [thick] (4.9,0.2) to [out=70,in=240] (5.1,0.4);
\end{tikzpicture}
\caption{The domains and the relevant graph approximations.}
\end{figure}
\end{center}
\vspace{-10pt}
While it seems natural to assume that the geometry induced by such graphs is random and will give rise to 'chaotic' behavior, that claim has to be tested. 
Therefore, in all what follows, data is given for both computations carried out using these types of graphs as well as classical random wave approximations.

\section{Longuet-Higgins' prediction and the number of local extrema} 
This section describes results obtained using the random wave model on $\mathbb{T}^2$, random linear combinations of spherical harmonics on
$\mathbb{S}^2$ and the grid graphs with randomly added edges.

\subsection{The torus $\mathbb{T}^2$.} Our approach was to use a random linear combination (with coefficients uniformly chosen from $[0,1]$) of explicit eigenfunctions on the torus. 
The table shows the eigenvalue, the dimension of the corresponding eigenspace, and the rescaled mean and variance written as a multiple of $k$, where everything is given in terms
of the $k-$th eigenfunction.
\vspace{-10pt}
\begin{center}
\begin{table}[h!]
\begin{tabular}{ l c c c c | l c c c c}
eigenvalue & dim  & mean & variance & eigenvalue & dim & mean & variance \\ [0.1cm] 
$1105\pi^2$ & 8  & 0.581$k$ & 0.137$k$ &		\,\,\,$2210 \pi^2$ & 8  & 0.589$k$ & 0.236$k$\\ [0.1cm] 
$1325 \pi^2$ & 6  & 0.574$k$ & 0.404$k$&		\,\,\,$3625 \pi^2$ & 8  & 0.583$k$ & 0.277$k$\\ [0.1cm] 
$2125 \pi^2$ & 8  & 0.582$k$ & 0.487$k$&		\,\,\,$5525 \pi^2$ & 12  & 0.582$k$ & 0.261$k$\\ [0.1cm] 
\end{tabular}
\caption{For each eigenvalue, the dimension of the eigenspaces and the mean and variance of our normalized number of extrema for $n=100$ random samples.}
\end{table}
\end{center}
\vspace{-45pt}

\subsection{The sphere $\mathbb{S}^2$.} We can use the explicit representation
of the spherical harmonics for additional testing: the multiplicity of the $\ell$-th
eigenvalues is 2$\ell$-1. We encounter two competing factors:
\begin{itemize}
\item the randomly generated functions should be even closer to chaotic eigenfunctions
as the eigenspace is very large but
\item the large eigenspace means that the $k-$th eigenvalue is going to be piecewise
constant and then jump for small $k$; the error term in the Weyl asymptotic is sharp
in the case.
\end{itemize}
 This has two effects, the first of which is helping our efforts because things get very
chaotic very quickly while the second effect is perturbing things for small $k$: there is
no unique '$k-$th eigenvalue' because the $i-$th eigenfunctions with 
$$i = \ell^2, \ell^2+1, \dots, \ell^2 + 2\ell + 1$$
all have the same eigenvalue. 

\begin{center}
\begin{table}[h!]
\begin{tabular}{ l  c c c }
 $\ell^2$ & 100 & 196 & 289 \\ [0.1cm] 
mean & 0.604$k$ & 0.606$k$ & 0.609$k$ \\ [0.1cm] 
variance & 0.608$k$ & 0.540$k$ & 0.368$k$ \\ [0.1cm] 
renormalized mean & 0.548$k$ & 0.564$k$ & 0.574$k$ \\ [0.1cm] 
\end{tabular}
\caption{Mean and variance for the number of local extrema of a linear combination of 
the $k-$th eigenfunctions for $k = \ell^2$ and $n=50$.}
\end{table}
\end{center}
\vspace{-30pt}
This data seems to suggest that the universal constant should be $\sim 0.6$, however, this
is misleading: the data was computed based on the assumption that $\ell^2$ is the $k-$th
eigenvalue. A natural renormalization is to interpret the $2\ell - 1$ dimensional eigenspace associated to the
eigenvalue $\ell^2$ in such a way that the 'average' value of $k$ should not be $k = \ell^2$
but rather
$$ k \sim \ell^2 + \ell + \frac{1}{2}.$$
Note that the correction tends to $1$ as $k$ becomes large, i.e. we expect both the normalized
und unnormalized mean to be slightly smaller than suggested by the data here.

\subsection{Random graphs.} Here we computed eigenfunctions on 100 random graphs
in each case. A minor difficulty occurs: our phrasing in terms of the Graph Laplacian uses implicitely Neumann conditions whereas  our calibration used Dirichlet conditions; this leads to certain boundary effects. 
We circumvented the problem by only looking at the number of extrema in the fixed 
box $[0.05, 0.95]^2$ and rescaling (thereby exploiting another universality phenomenon:
quantum chaos in a subregion should not behave any differently than in the whole space). The results are summarized in the following table.
\begin{center}
\begin{table}[h!]
\begin{tabular}{ l  c c c }
 $k$ & 100 & 200 & 300 \\ [0.1cm] 
5 random edges & (0.69$k$, 0.06$k$) & (0.52$k$, 0.07$k$) & (0.59$k$, 0.71$k$) \\ [0.1cm] 
10 random edges &  (0.71$k$, 0.06$k$) &  (0.53$k$, 0.22$k$)  & (0.59$k$, 0.37$k$)\\ [0.1cm] 
20 random edges & (0.75$k$, 0.16$k$) & (0.56$k$, 0.13$k$) &  (0.60$k$, 0.28$k$)\\ [0.1cm] 
\end{tabular}
\vspace{5pt}
\caption{Mean and variance for the number of local extrema of the $k-$th eigenfunction
on a $150 \times 150$ grid graph for a different number of random edges added. For each
case mean/variance was taken over a random sample with $n=100$.}
\end{table}
\end{center}
\vspace{-35pt}

\section{Discrepancy of Local Extrema}
\subsection{Discrepancy} This section is devoted to the introduction of the notion of discrepancy;
it is a widely used and foundational concept in the theory of quasi-Monte Carlo methods, where the distribution
properties of a set of points directly imply error bounds for numerical integration schemes using these
set of points. We refer to the classical books of Niederreiter \cite{nied} and Drmota \& Tichy \cite{drm} as well as the recent monograph of Dick \& Pillichshammer \cite{dick} for more detailed information about discrepancy.
Formally, let $\mathcal{P} = \left\{p_1, \dots, p_N\right\}$ be
a set of points in the unit square, i.e. $\mathcal{P} \subset [0,1]^2$ and write $p_i = (x_i, y_i)$.
We define the \textit{star-discrepancy} $D_N^*$ via
$$ D_N^*(\mathcal{P}) := \sup_{0 \leq x,y \leq 1}{\left| \frac{\# \left\{1\leq i \leq N: 0 \leq x_i \leq x \wedge 0 \leq y_i \leq y\right\}}{|\mathcal{P}|} - xy\right|}.$$
A graphical description is as follows: given the set of all axis-parallel rectangles anchored in $(0,0)$,
we measure the maximal deviation between the area of the rectangle and the relative size of the set
$\mathcal{P}$ contained in the rectangle. If one wishes to replace the set of rectangles anchored in
the origin by the set of all axis-parallel rectangles, this gives rise to the notion of $\textit{discrepancy}$.
We will only work with star-discrepancy since it is technically easier and $D_N^* \leq D_N \leq
2D_N^*$. A celebrated result of Schmidt \cite{schm} states that
$$ D_N(\mathcal{P}) \geq \frac{1}{100}\frac{\log{N}}{N}$$
for any two-dimensional set $\mathcal{P}$ with $\# \mathcal{P} = N$ and this is asymptotically best possible. 

\begin{center}
\begin{figure}[h!]
\begin{tikzpicture}[scale=3]
  \draw[thick] (0,0) -- (1,0);
  \draw[thick] (0,0) -- (0,1);
  \draw[thick, dashed] (0,0.34) -- (0.74,0.34);
  \draw[thick, dashed] (0.74,0) -- (0.74,0.34);
\draw [fill, ultra thick] (0.1,0.9) circle [radius=0.01];
\draw [fill, ultra thick] (0.2,0.3) circle [radius=0.01];
\draw [fill, ultra thick] (0.3,0.7) circle [radius=0.01];
\draw [fill, ultra thick] (0.4,0.1) circle [radius=0.01];
\draw [fill, ultra thick] (0.5,0.5) circle [radius=0.01];
\draw [fill, ultra thick] (0.6,0.9) circle [radius=0.01];
\draw [fill, ultra thick] (0.7,0.2) circle [radius=0.01];
\draw [fill, ultra thick] (0.8,0.5) circle [radius=0.01];
\draw [fill, ultra thick] (0.9,0.7) circle [radius=0.01];

\draw[thick] (2,0) -- (3,0);
\draw[thick] (2,0) -- (2,1);
  \draw[thick, dashed] (2,0.96) -- (2.3,0.96);
  \draw[thick, dashed] (2.3,0) -- (2.3,0.96);
\draw [fill, ultra thick] (2.23,0.23) circle [radius=0.01];
\draw [fill, ultra thick] (2.23,0.56) circle [radius=0.01];
\draw [fill, ultra thick] (2.23,0.89) circle [radius=0.01];
\draw [fill, ultra thick] (2.56,0.23) circle [radius=0.01];
\draw [fill, ultra thick] (2.56,0.56) circle [radius=0.01];
\draw [fill, ultra thick] (2.56,0.89) circle [radius=0.01];
\draw [fill, ultra thick] (2.89,0.23) circle [radius=0.01];
\draw [fill, ultra thick] (2.89,0.56) circle [radius=0.01];
\draw [fill, ultra thick] (2.89,0.89) circle [radius=0.01];
\end{tikzpicture}
\caption{Set of points and a rectangle containing three points.}
\end{figure}
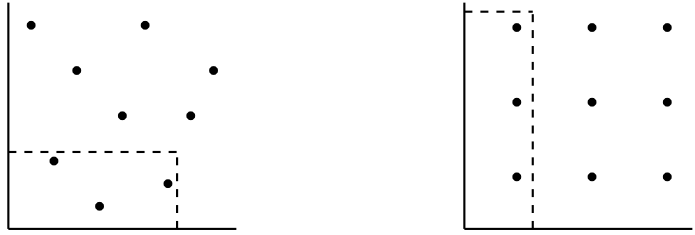
\end{center}

A simple calculation shows that a regular grid always satisfies $D_N($grid$)\gtrsim N^{-1/2}$. 
Using the definition, it follows that for a set of $N$ points $\mathcal{P}$ with discrepancy
$D_N^*(\mathcal{P})$ and any axis-parallel rectangle $R \subset [0,1]^2$, we have 
$$ |R| -2D_N^*(\mathcal{P}) \leq \frac{\# (\mathcal{P} \cap R)}{\# \mathcal{P}} \leq |R| + 2D_N^*(\mathcal{P}).$$

\subsection{Sets with small discrepancy}
In this section we briefly describe the different constructions of deterministic point sets, which serve as benchmark with which we will compare the regularity of the local extrema of Laplacian eigenfunctions --
we will create quantum chaos using again random linear combinations of eigenfunctions on the
torus $\mathbb{T}^2$ as well as eigenfunctions on random graphs. The sphere $\mathbb{S}^2$
is not as suited because no isometric embedding $\phi:\mathbb{S}^2 \rightarrow [0,1]^2$ exists
(one could study other notions of discrepancy on $\mathbb{S}^2$ but this outside of the scope of
this paper).

\subsubsection{Random points} The star discrepancy of a point set with $N$ random elements in $[0,1]^2$ is a random
variable with expectation $D_N^* \sim N^{-1/2}.$ We will always include discrepancy results for random point as a natural benchmark of the simplest possible method.

\subsubsection{Regular grid}
In contrast to random point sets, the regular grid 
$$ \mathcal{G}_m=\left\{ \left( \frac{i}{m}, \frac{j}{m} \right) : 0 \leq i, j \leq m-1\right\} \qquad
\mbox{of}~N=m^2~\mbox{points}$$
is a highly structured point set with star discrepancy 
$$D_N(\mathcal{G}_m)= \frac{2m-1}{m^2} \sim \frac{2}{\sqrt{N}}.$$ 
It is easy to see that shifting the grid by a random vector $x \in [0,1/m]^2$ can only decrease the value of the star discrepancy; however, at most by a constant factor of $1/2$.

\subsubsection{Hammersley point sets} Hammersley point sets and their generalizations are classical examples of low discrepancy point sets created using number theory. Let $n=\sum_{i=0}^{\infty} n_i b^{i}$ be the $b$-adic expansion of the integer $n$ with $0\leq n_i \leq b-1$. Then the \emph{radical-inverse function}, $\phi_b: \NN \rightarrow [0,1]$, is defined as
$$ \phi_b(n) = \frac{n_0}{b} + \frac{n_1}{b^2} + \ldots = \sum_{i=0}^{\infty} \frac{n_i}{b^{i+1}}.$$
Furthermore, for $N=b^m$ the \emph{two-dimensional Hammersley point set in base $b$} is then defined by
$$ \mathcal{H}_{b,m} = \left \{ \left ( \phi_b(n), \frac{n}{b^m} \right): 0 \leq n \leq b^m-1 \right \}. $$
We refer to \cite{fau} for an explicit formula for $D_N^*(\mathcal{H}_{b,m})$ and its generalization using permutations of the digits in the $b$-adic expansion of $n$.

\subsection{Results.}\label{disc} Our results are summarized in the following table. Each entry gives
both the mean as well as the standard deviation of the discrepancy sampled over $n=20$ 
random constructions (which is already sufficient for a conclusive result as the distribution
is tightly concentrated around its mean, note the small variance). We refer to the appendix 
for additional technical details.

\begin{center}
\begin{table}[h!]
\begin{tabular}{ l | c c c c }
$\#$ points & 50 & 100 & 200 & 500 \\ 
\hline  
\textsc{random-graph}	& 	(0.110, 0.014) 	& (0.074, 0.012)	& (0.058, 0.011)	& (0.029, 0.003)\\ [0.1cm] 
\textsc{random-torus} 	& 	(0.105, 0.005) 	& (0.078, 0.003) 	& (0.055, 0.004) 	& (0.033, 0.004) \\ [0.1cm] 
random points  			& 	(0.202, 0.046)		& (0.116, 0.019)	& (0.088, 0.016)	& (0.054, 0.011)	\\ [0.1cm] 
standard grid  			& 0.265				& 0.190			& 0.128			& 0.085			\\ [0.1cm] 
shifted grid  				& 0.133				& 0.095			& 0.064			& 0.043		\\[0.1cm] 
Hammersley sets 			& 0.082				& 0.0544			& 0.035			& 0.022	\\[0.1cm] 
\end{tabular}
\vspace{10pt}
\caption{Mean and standard deviation of the discrepancy of various sets.}
\end{table}
\end{center}

We note that the discrepancy is clearly decaying implying that the distribution of local 
extrema tends towards uniform distribution (as one would expect). However, it is certainly
surprising that the discrepancy of the local extrema seems to be smaller than both that
of the standard/shifted grid as well as that of fully random points. It seems natural to conjecture
that asymptotically the discrepancy of
$$\mbox{the set}~\mathcal{P}_k~\mbox{of local extrema of the}~k-\mbox{eigenfunction should scale as}~~
\sim \frac{c}{\sqrt{k}}.$$

\section{Methodology and technical details}

\subsection{Detection of local extrema} Finding local extrema computationally is in general difficult: any function
might have tiny oscillations at a small scale which could potentially never be discovered by a computer, or, reversely, inaccuracies of the computer might generate tiny artificial extrema depending on the geometry and parameters of the underlying grid. We start by giving some heuristics why this is not to be expected in the case of Laplacian eigenfunctions: basically, eigenfunctions of second order partial differential equations of \textit{elliptic} type with constant coefficients come equipped with a natural length scale on which oscillations occur. This insight motivates to count extrema using a local, fast and easy-to-implement algorithm, which requires the evaluation of the function $f$ on a fine grid imposed on $[0,1]^2$. 
Despite our heuristics that ensure the accuracy of the local extrema count, we also sketch an alternative, topological algorithm. While this algorithm is computationally more demanding, it is provably reliable in much more general situations than the one we study here.

\subsubsection{Stability of the problem.} A standard estimate for a Laplacian eigenfunction satisfying
$$ -\Delta u = \lambda u \qquad \mbox{on some domain with Dirichlet conditions at the boundary}$$
is that
$$ \|\nabla u\|_{L^{\infty}} \sim \lambda^{\frac{1}{2}}\|u\|_{L^{\infty}}.$$
This clarifies a natural heuristic saying that such an eigenfunction may be seen as a collection of
waves with wavelength $\lambda^{-1/2}$. Furthermore, the elliptic equation $-\Delta u = \lambda u$
implies quantitative control on the concavity of $u$ around a local maximum (and, conversely, quantitative
control on the convexity around a local minimum). This quantitative control weakens whenever $u$ has a local extremum with $|u|$ being very small -- something that is not to be expected in the case of quantum chaos. This implies that, at least in the generic case, two extrema are at least distance $\sim \lambda^{-1/2}$ away from each other.

\subsubsection{Local Method}
The local method is straightforward and easy to implement as it simply checks all eight neighbors of a given grid point $(i,j)$. If $f(i,j)$ is greater (resp. smaller) than the value of every neighbor the algorithm reports a local extremum (resp. minimum).
\begin{center}
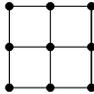
\begin{figure}[h]
\begin{tikzpicture}[scale=2.7]

\draw [fill, ultra thick] (1.7,0) circle [radius=0.01];
\draw [fill, ultra thick] (1.9,0) circle [radius=0.01];
\draw [fill, ultra thick] (2.1,0) circle [radius=0.01];

\draw [fill, ultra thick] (1.7,0.2) circle [radius=0.01];
\draw [fill, ultra thick] (1.9,0.2) circle [radius=0.01];
\draw [fill, ultra thick] (2.1,0.2) circle [radius=0.01];

\draw [fill, ultra thick] (1.7,0.4) circle [radius=0.01];
\draw [fill, ultra thick] (1.9,0.4) circle [radius=0.01];
\draw [fill, ultra thick] (2.1,0.4) circle [radius=0.01];

  \draw (1.7,0) -- (2.1,0);
  \draw (1.7,0.2) -- (2.1,0.2);
  \draw (1.7,0.4) -- (2.1,0.4);
  \draw (1.7,0) -- (1.7,0.4);
  \draw (1.9,0) -- (1.9,0.4);
  \draw (2.1,0) -- (2.1,0.4);
  \draw (2.1,0) -- (2.1,0.4);

\end{tikzpicture}
\caption{Checking 8 adjacent grid points.} \label{pic7}
\end{figure}
\vspace{-20pt}
\end{center}
Choosing the appropriate grid size is of course essential. 
Our rule of thumb was that the grid size $h$ should scale as $h \sim 0.2 \lambda^{-1/2}$. This implies
that the grid is really smaller than the wavelength, certainly a necessary condition for the approximation to
work. The situation simplifies a bit since we have the explicit form of the eigenfunctions at our disposal. For example on
$\mathbb{T}$, the eigenfunctions associated to the eigenvalues $2210\pi^2$ with the highest oscillation frequency are $\sin{(\pi x)}\sin{(47 \pi y)}$ and $\sin{(\pi y)}\sin{(47 \pi x)}$. This implies that a 
$150 \times 150$ grid certainly satisfies the rule of thumb. 
We additionally checked the stability of the method by performing iterative refinements on larger grids and comparing the results to ensure sufficient stability.

\subsubsection{Topological method} 
The topological method uses the concept of \emph{persistent homology} and is a standard method in computational topology. We refer to the book of Edelsbrunner \& Harer \cite{edels} for further details.
We evaluate the function again on a finite number of points (not necessarily a regular grid), which we extend to a simplicial complex. We then compute the lower (upper) star filtration of this complex and use it to compute the \emph{$0$-th extended persistence diagram} of $f$. This diagram describes the evolution of the $0$-dimensional homology of the sequence of sub- and superlevel sets of $f$ from which we can read off the number of extrema. While this method is more involved and computationally more expensive, its main advantage lies in the various stability results of persistent topology \cite{stab1, stab2, edels} which formally ensure that this method indeed detects all relevant extrema of $f$ on any scale.
We used a standard implementation of the persistence algorithm to confirm the results of the local algorithm.

\subsection{Discrepancy} The purpose of this section is to describe various technical details regarding the results presented
in Section \ref{disc}. One common necessity is that in our construction of random Laplacian eigenfunctions
it is not a priori clear how many local extrema one can expect; however, due to the underlying
universality phenomenon, it was rather easy to pick the right $k$ such that the $k-$th eigenfunction
will have 'on average' the desired number of local extrema (indeed, this is how we discovered the
universal behavior for the number of local extrema).

\subsubsection{Computation} For the computation of the discrepancy we use a recent implementation of the Dobkin-Eppstein-Mitchell algorithm \cite{dob} by Magnus Wahlstr\"{o}m \cite{wahl}, which is freely available online. This algorithm computes the star discrepancy exactly; for details on the implementation and a fast randomized algorithm for the approximation of the star discrepancy of large multi-dimensional point sets we refer to \cite{doerr}.

\subsubsection{Random graphs}
We used a $n \times n$ grid graph, where $n = 120$ for eigenfunctions containing up to 200 extrema and $n = 170$ for higher accuracy in the final case. Furthermore, we always added 10 edges (in the first three
cases) and 50 random edges (in the final case), respectively, between edges chosen fully at random (the precise number of edges added seemed to have very little impact on the overall results). We then looked for the $k$ such that the the $k$th eigenfunction has, on average, the desired number of local extrema: on the $170 \times 170$ graph, this was, for example, the 1050th eigenfunction. However, the results seem very stable and it does not seem to have a big effect whether one chooses the 1050th eigenfunction or, say, the 1037th. Since we are again computing Neumann solutions with the usual boundary phenomena, we again did an intersection of the points with the fixed box $[0.05, 0.95]^2$ followed by the appropriate rescaling.

\subsubsection{Random combinations on the torus.} The problem that arose here is that, at least for small
eigenvalues, not many of them have high multiplicity; it is thus not always possible to produce random
combinations having on average the desired number of local extrema. Our choice was as follows.
\begin{center}
\begin{tabular}{ l c c  }
eigenvalue & dim(eigenspace) & expected number of local extrema \\
$125\pi^2$ & 4 & $\sim$ 55  \\
$221\pi^2$ & 4 & $\sim$ 100  \\
$481\pi^2$ & 4 & $\sim$ 220  \\
$1105\pi^2$ & 8 & $\sim$ 500  \\
\end{tabular}
\end{center}

In particular, it should be noted that in the following comparison in the cases of 50 and 200 points,
the point sets arising as local extrema will have $10\%$ more points than the sets used for benchmark
(which, however, will changes the resulting data by less than $10\%$ due to superlinear scaling).

\subsubsection{Deterministic sets}
The standard grid, the randomly shifted grid as well as fully random points require no further explanation;
regarding parameters, the grid was computed with $m=7, 10, 15, 23$ to generate point sets with $m^2$ many points ($m = 15$ was chosen instead of $m=14$ to compensate for the corresponding effect on the torus). Similarly, the same numbers were used as basis for the generation of corresponding Hammersley point sets of the same size.

\bigskip
\textbf{Acknowledgment.} The authors thank Olga Symonova and Michael Kerber for sharing their implementation of the persistence algorithm. F.P. was supported by the Graduate School of IST Austria. S.S. was partially supported by SFB 1060 of the DFG.

\end{document}